\documentclass[twocolumn,showpacs,preprintnumbers,amsmath,amssymb]{revtex4}

\usepackage{graphicx}
\usepackage{amssymb}
\usepackage{hyperref}

\begin{document}
\title{Structural, magnetic and thermal properties of the substitution series
Ce$_2$(Pd$_{1-x}$Ni$_x$)$_2$Sn}

\author{J.G. Sereni $^{1 *}$, G. Schmerber $^2$, A. Braghta $^3$,  B.
Chevalier $^4$ and J.P. Kappler $^2$}
\address{$^1$ Divisi\'on Bajas Temperaturas (CAB - CNEA), Conicet, 8400 S.C.
Bariloche, Argentina\\
$^2$ IPCMS, UMR 7504 CNRS-ULP, 23 rue de Loess, B.P. 43 Strasbourg
Cedex 2, France\\
$^3$ Département de Physique, Universit\'e de Guelma, 24000
Guelma,
Algeria\\
$^4$ CNRS, Universit\'e de Bordeaux, ICMCB, 87 av. Dr. Schweitzer,
33608 Pessac Cedex, France}
\date{\today}

\begin{abstract}

{Structural, magnetization and heat capacity studies were
performed on Ce$_2$(Pd$_{1-x}$Ni$_x$)$_2$Sn ($0 \leq x \leq 1$)
alloys. The substitution of Pd atoms by isoelectronic Ni leads to
a change in the crystallographic structure from tetragonal (for $x
\leq 0.3$) to centered orthorhombic lattice (for $x \geq 0.4$).
The volume contraction thorough the series is comparable to the
expected from the atomic size ratio between transition metal
components. The consequent weak increase of the Kondo temperature
drives the two transitions observed in Ce$_2$Pd$_2$Sn to merge at
$x = 0.25$. After about a 1\% of volume collapse at the structural
modification, the system behaves as a weakly magnetic heavy
fermion with an enhanced degenerate ground state. Notably, an
incipient magnetic transition arises on the Ni-rich size. This
unexpected behavior is discussed in terms of an enhancement of the
density of states driven by the increase of the $4f$-conduction
band hybridization and the incipient contribution of the first
excited crystal field doublet on the ground state properties.}

\vspace{0.5cm}

$^*$ E-mail-address of corresponding author:
jsereni@cab.cnea.gov.ar

\end{abstract}

 \maketitle

%\keywords{Keywords: Cerium Intermetallics, Critical Points,
%Magnetic Phase Diagrams} \end{frontmatter}

\section{Introduction}

The large family of Rare Earth based ternary compounds with the
formula R$_{2}$T$_{2}$X (where R=Ln \cite{Hulliger95,Giovannini98}
and Ac \cite{Miranblet93}, T=Transition Metals of the VIII group
\cite{Hulliger95,Gordon95,Kacharovsky} and X=early $p$-metals
\cite{Gordon95}) have been investigated over the past two decades
because of the variety of their magnetic behaviors. Their
tetragonal Mo$_{2}$FeB$_{2}$-type structure \cite{Peron93} is
strongly anisotropic and can be described as successive `T+X' (at
z=0) and `R' (at z=1/2) layers. Within the Ln series, Ce and Yb
elements have shown some outstanding features like double magnetic
transitions in Ce$_2$Pd$_2$Sn \cite{Fourgeot96}, reentrance of
magnetic order in the Yb$_{2}$Pd$_{2}$(In,Sn) system
\cite{Bauer05} and under applied pressure \cite{Bauer09}, a gap in
the solid solution in Ce$_{2\pm x}$Pd$_{2\pm y}$In$_{1\pm z}$
alloys \cite{Giovannini98} with ferromagnetic FM and
antiferromagnetic AF behaviors in the respective Ce-rich and
Pd-rich branches. In the case of Yb$_2$Pt$_2$Pb \cite{Kim} and
Ce$_2$Pd$_2$Sn \cite{Laffarge96}, the formation of
Shastry-Sutherland lattices \cite{Miyahara,Ce2Pd2Sn} were reported
as the result of geometrical frustration constrains originated in
a triangular network of magnetic atoms. In the latter compound,
which shows two magnetic transitions, the Shastry-Sutherland
lattice formed at $T_{N}=4.7$\,K is overcome by a FM ground state
GS at $T_{C}=3.0$\,K in a first order transition \cite{Ce2Pd2Sn}.
Further studies under applied magnetic field demonstrated that
such an exotic phase can be suppressed by the application of a
moderate magnetic field $H_{cr}= 0.11$\,T \cite{PRBfield}

It is well known that the magnetic behavior of most of Ce-based
compounds which adopt magnetic GS is governed by a competition
between Kondo and Ruderman-Kittel-Kasuya-Yosida RKKY magnetic
interactions. Since the Kondo effect is due to the local screening
of $4f$ moments by conduction-electron spins through an AF
coupling exchange parameter $J_{ex}$ and RKKY also depends on that
coupling parameter for polarizing the conduction spins, both
mechanisms are basically driven by the same parameter.
Experimentally, $J_{ex}$ can be tuned by two control parameters:
pressure and chemical potential variation.

Since the experimental results confirm Ce$_2$Pd$_2$Sn as one of
the scarce examples of FM-GS among Ce intermetallics, with stable
magnetic moments and a weak Kondo effect, to drive the magnetic
transition by increasing the hybridization effect (i.e. increasing
the $J_{ex}$ strength) can provide valuable information concerning
the stability of the mentioned Shastry-Sutherland intermediate
phase. Structural pressure, produced by the substitution of larger
size Pd by smaller Ni atoms, is one of the experimental
possibilities as already proved in the study of CePd$_{1-x}$Ni$_x$
binary alloys \cite{CePdNi}.

In the case of Ce$_{2}$(Pd$_{1-x}$Ni$_x$)$_2$Sn alloys, it is
known that the isotypic Ce$_2$Ni$_2$Sn compound forms with a
centered orthorhombic structure of W$_2$CoB$_2$ type
\cite{DiSalvo95}, which behaves as a magnetically ordered Kondo
system \cite{Fourgeot95}. However, to our knowledge no detailed
study of the structural transition and its consequence on the low
temperature magnetic properties was carried up to now like in
Ce$_{2}$(Pd$_{1-x}$Ni$_x$)$_2$In \cite{DiSalvo96}

In this work we present the results obtained from the study of the
magnetic properties of the Ce$_{2}$(Pd$_{1-x}$Ni$_x$)$_2$Sn alloys
in order to establish the range of stability of both
Mo$_{2}$FeB$_{2}$ and W$_{2}$CoB$_{2}$-type structures under the
effect of structural pressure provided by the substitution of Pd
atoms by smaller isoelectronic Ni ones.

\section{Experimental Details}

Polycrystalline Ce$_{2}$(Pd$_{1-x}$Ni$_x$)$_2$Sn and La isotypic
samples were prepared by conventional tri-arc melting of
appropriate amounts of 99.99\% Ce, La, Pd and Sn, and 99.999\%
pure Ni, under flowing purified argon atmosphere on a water-cooled
copper hearth. The buttons were remelted several times to ensure
good homogeneity. The weight losses after arc melting were less
than 0.2 wt.\% of total mass (each sample of a total amount of ca.
1g). Each alloy was wrapped in Ta-foil, sealed in high evacuated
silica tube and heat-treated for 3 weeks at 750$^o$C. This
annealing leads to the disappearance of some extra peaks and less
unfolding. The alloys show a metallic-gray lustre and were air
sensitive and pyrophoric. The crystallographic structure of the
annealed alloys were collected at room temperature employing a
Bruker AXS D8 Advance X-ray diffractometer with monochromatic
CuK$_{\alpha 1}$ radiation, within $15^o < 2\theta < 80^o$, and
settings of 55\,kV and 25\,mA. The system was equipped with a
Sol'X detector. The surface of the samples were mechanically
cleaned by scraping slightly with a diamond file in a gloved box
under argon gas flow. After this cleaning procedure the samples
were crushed slowly under C$_6$H$_{12}$ with average grain size
$\leq 5 \mu m$. The peak positions and major peak intensities are
all consistent with those expected for materials in the
Mo$_2$FeB$_2$-type tetragonal structure for $x < 0.3$ and the
W$_2$CoB$_2$ orthorhombic structure for $x >0.4$.

The susceptibility and magnetization measurements were performed
using a Superconducting Quantum Interference Device (SQUID)
magnetometer running between 1.8\,K and room temperature, and with
applied magnetic fields up to 5T. Specific heat was measured
between 1.5 and 40\,K using a standard heat pulse method with
$\Delta T/T \approx 1\%$.

\section{Experimental Results}

\begin{figure}
\begin{center}
\includegraphics[angle=0,width= 0.5 \textwidth] {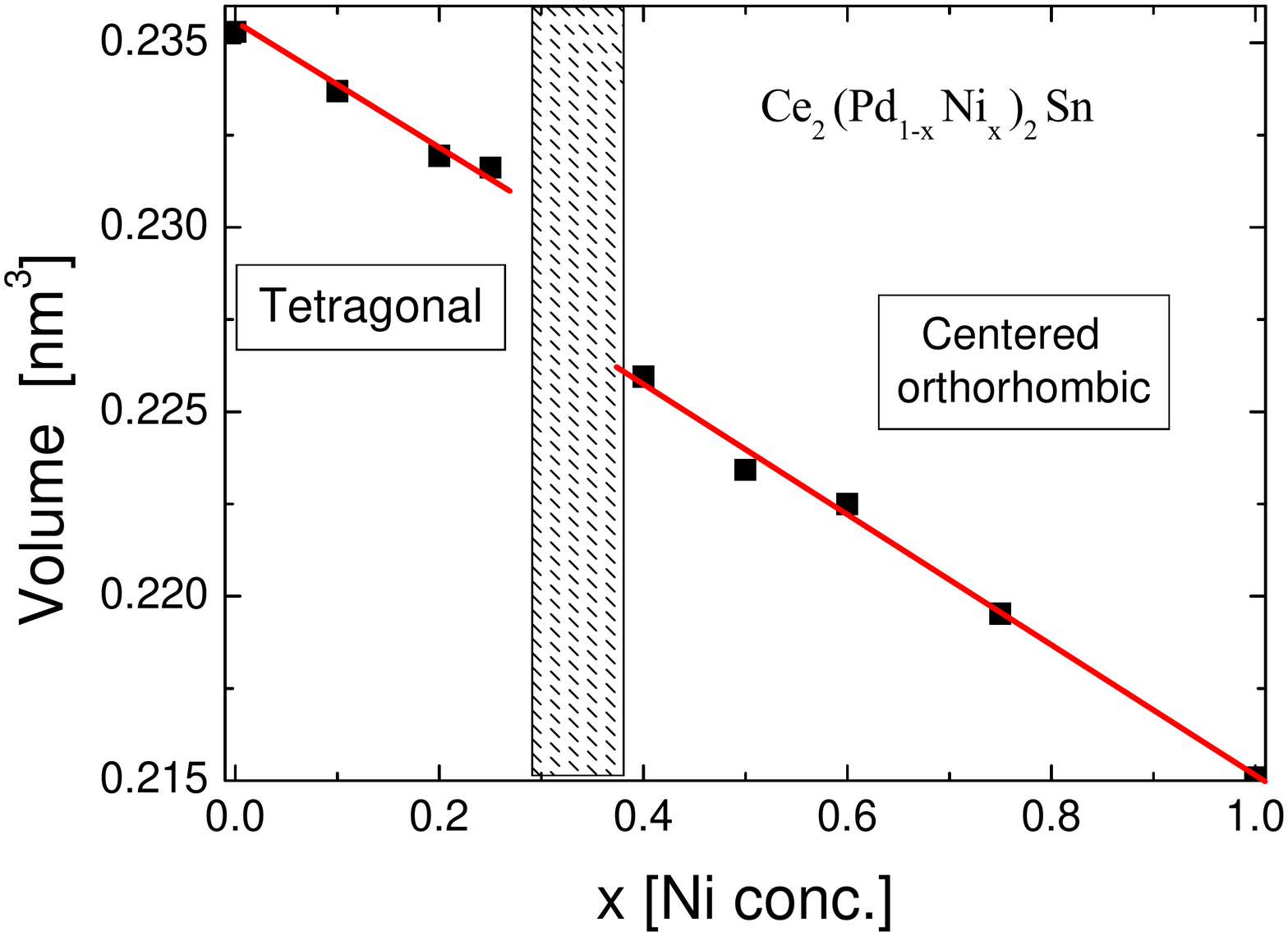}
%{/home/jsereni/papers/publicaz02/pdnial3/textos/F1latpar}
\end{center}
\caption{(Color online) Unit cell volume dependence on Ni content,
showing a discontinuity at the change of structure.} \label{F1}
\end{figure}

\subsection{Crystalline structure}
Despite of their different crystalline structures, Ce$_2$Pd$_2$Sn
and Ce$_2$Ni$_2$Sn practically form a continuous solid solution.
These stannide alloys crystallize in tetragonal Mo$_2$FeB$_2$-type
structure for $x<0.3$ and orthorhombic W$_2$CoB$_2$-type for $x
> 0.4$, with a small gap between $0.3<x<0.4$. The substitution of Ni by Pd
in Ce$_2$(Pd$_{1-x}$Ni$_x$)$_2$Sn produces an overall volume
contraction of about 8.6\%, including the volume collapse of about
1\% at the crystallographic transition between both structures
(see Fig.~\ref{F1}). Such a volume reduction between both
stoichiometric extremes is comparable to the contraction evaluated
taking into account the relative atomic volume $V$ difference
between Pd and Ni, i.e. 100 *($V_{Pd}-V_{Ni})/V_{Pd}\approx 26\%$.

On the Pd side (Mo$_2$FeB$_2$-type structure) the volume variation
stems from the change of the two lattice parameters (c.f. $a$ and
$c$), with the consequent variation of the Ce-Ce spacing. On the
Ni-rich side, the Ce-Ce spacing decreases due to a $\approx 4\%$
decrease of $a$ (from 0.4562\,nm to 0.4381\,nm) and $\approx 2\%$
$b$ (from 0.5829\,nm to 0.5728\,nm), whereas the $c$ parameter
even increases by $\approx 1\%$ (from 0.8496\,nm to 0.8570\,nm).
In fact, from the neutron powder diffraction on Ce$_2$Pd$_2$Sn
\cite{Laffarge99,Laffarge99b}, one can deduce that the first
neighbors Ce-Ce spacing is along $c$ axis (Ce1-Ce2=0.4038\,nm
\cite{Fourgeot96} and Ce3-Ce4=0.5902\,nm).

\subsection{Magnetic Susceptibility}

\begin{figure}
\begin{center}
\includegraphics[angle=0,width= 0.45 \textwidth] {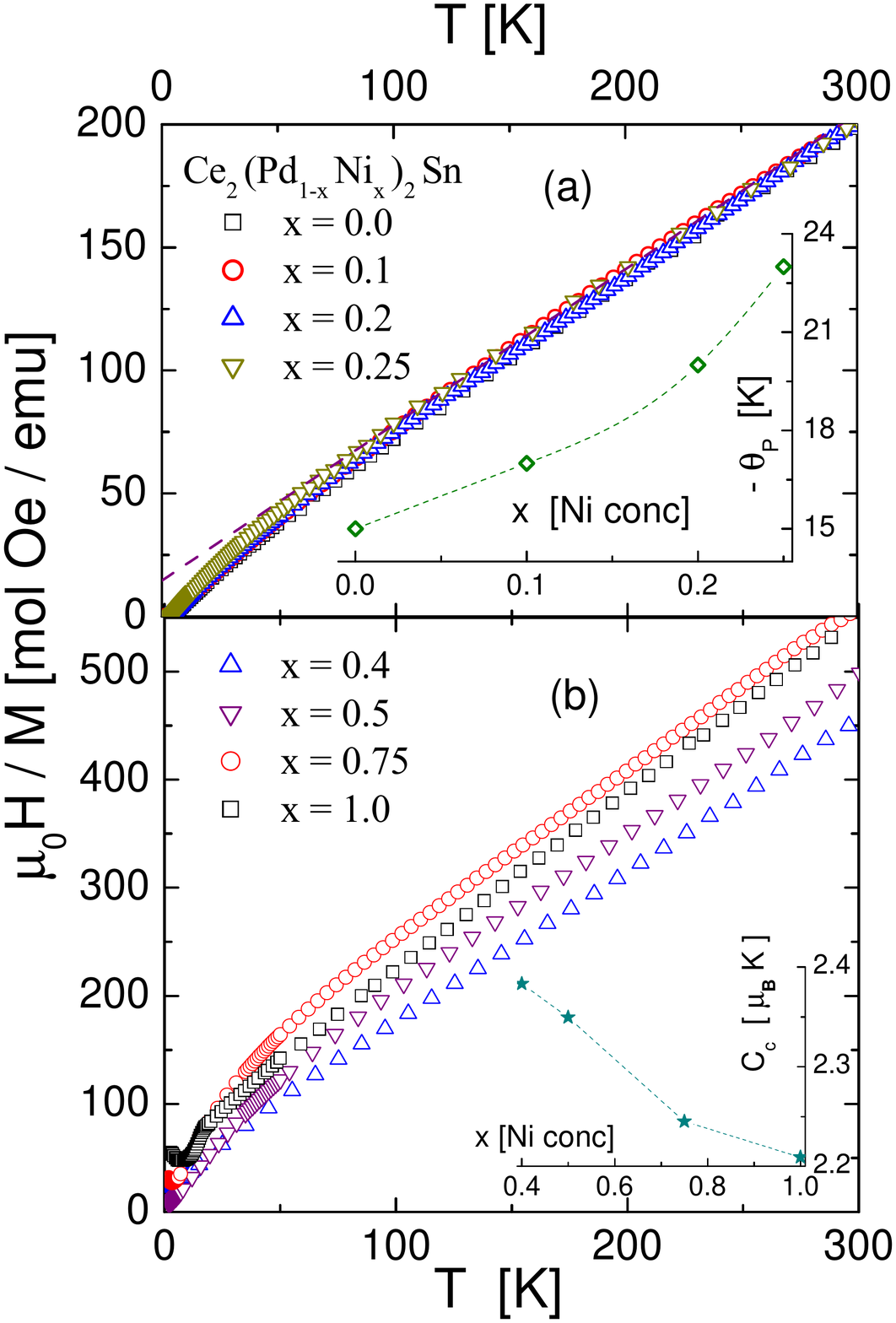}
%{/home/jsereni/papers/publicaz02/pdnial3/textos/F1latpar}
\end{center}
\caption{(Color online) High temperature inverse susceptibility
for (a) Pd-rich region and (b) Ni-rich one, measured with $\mu_0
H=1$\,kOe. Respective insets represent the Ni concentration
dependencies of (a) Curie-Wiess temperature and (b) Ce effective
magnetic moment $\mu_{eff}$ on the Ni-rich side.} \label{F2}
\end{figure}

The inverse of the magnetic susceptibility of
Ce$_2$(Pd$_{1-x}$Ni$_x$)$_2$Sn measured between 1.8\,K and room
temperature is shown in Figs.~\ref{F2}a and b. On the Pd-rich side
a Curie-Weiss behavior $1/\chi = (T+\theta_P)/C_c$ is observed for
$T>100$\,K, see dashed line in Fig.~\ref{F2}a. The Curie-Weiss
paramagnetic temperature $\theta_P$, extrapolated from $T>100$\,K,
slightly decreases from -15\,K at x = 0 to -23\,K at x = 0.25, see
the inset in Fig.~\ref{F2}. Coincidentally, the computed effective
magnetic moments, computed from the Curie constant $C_C$, are
$\mu_{eff} = 2.5\pm 0.02 \mu_B$, which corresponds to the
Ce$^{3+}$ magnetic moment. The negative curvature of $1/\chi(T)$
below 11\,K can be attributed to the electric crystal field effect
CFE as the excited levels become thermally depopulated.

A different behavior is observed in Fig.~\ref{F2}b for the Ni-rich
side. Using the same Curie-Weiss description, increasing values of
$\theta_P$ from -25\,K to $-55\pm 5$\,K are extracted. The Curie
constant decreases concomitantly with Ni concentration, from
$2.38\pm 0.02\mu_B$ for x = 0.4 to $2.2\mu_B$ for x = 1, as shown
in the inset of Fig.~\ref{F2}b.

%\subsubsection{Low temperature Magnetization}

\begin{figure}
\begin{center}
\includegraphics[angle=0,width= 0.45 \textwidth] {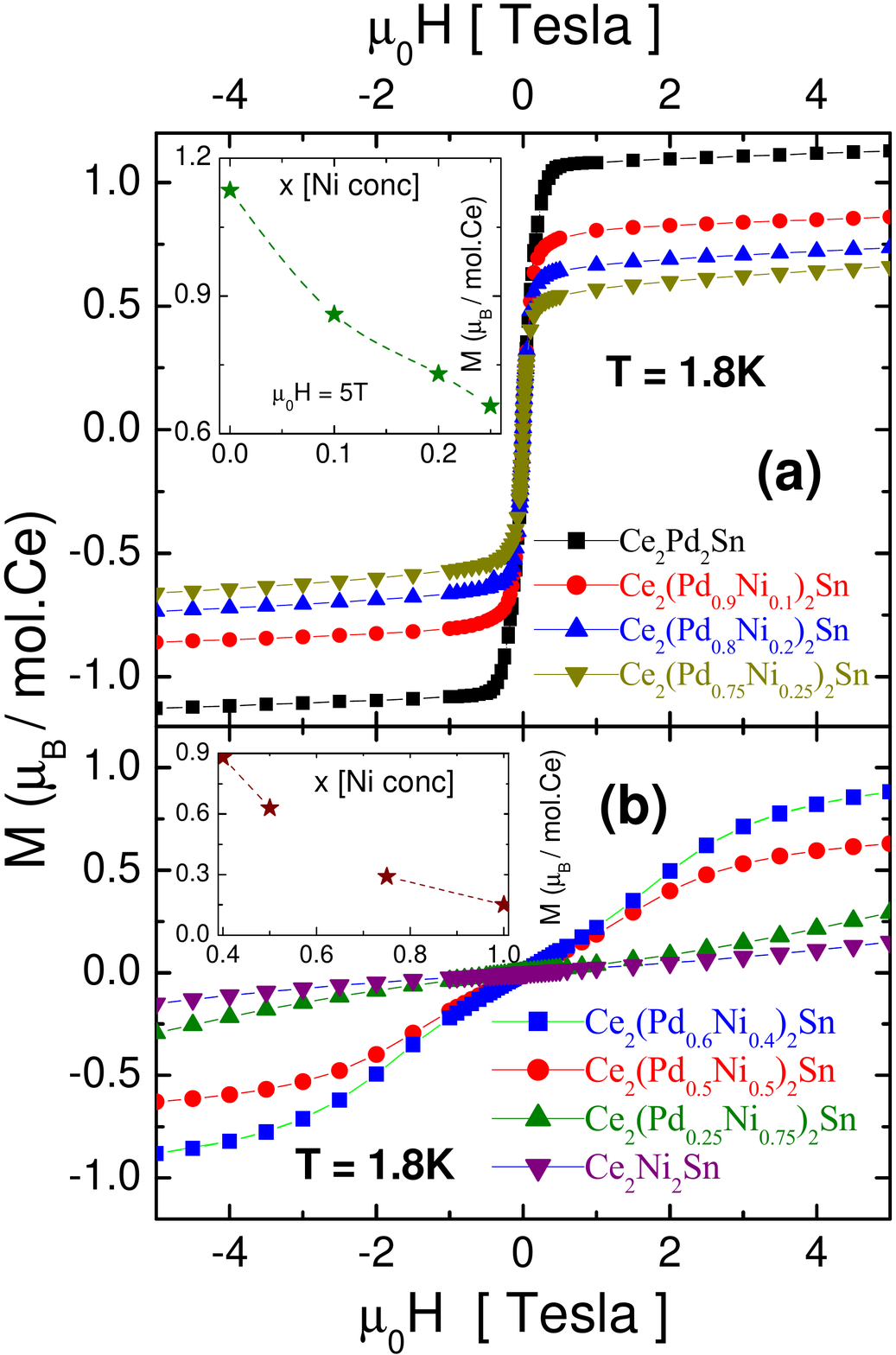}
%{/home/jsereni/papers/publicaz02/pdnial3/textos/F1latpar}
\end{center}
\caption{(Color online) Field dependent magnetization of the 1.8K
isotherms, (a) for the Pd-rich region and (b) for the Ni-rich on,
measured up to 5T. Respective insets represent the Ni
concentration dependencies of the magnetization values at $\mu_0 H
= 5$\,T.} \label{F3}
\end{figure}

The low temperature magnetization was measured at selected
temperatures ($T = 1.8$, 3, 4 and 5\,K) up to $\mu_0 H=5$\,Tesla.
The most relevant information is extracted from the $M(H)$
isotherms measured at $T=1.8$\,K which are displayed in
Figs.~\ref{F3}a and b for the respective Pd-rich and Ni-rich
sides. On the Pd-rich one, the $M(H)$ dependence shows the typical
behavior for a FM phase which saturates around 0.5\,T, but with a
decreasing moment from $1.13\mu_B$ for x = 0 to $0.66\mu_B$ for x
= 0.25. These values were measured at $\mu_0 H=5$\,T and are shown
in the inset of Fig.~\ref{F3}a. Despite the expected saturation
moment for the free Ce$^{3+}$ ion with total angular moment
$J=5/2$ according to Hund's rules is $g_J J = 2.14 \mu_B$, the
observed value is coherent with a doublet ground state GS with low
but increasing hybridization (i.e. Kondo screening) effect. The
measured value for the stoichiometric compound Ce$_2$Pd$_2$Sn is
in agreement with those reported in the literature
\cite{Laffarge99b,PRBfield}

Within the Ni-rich phase, the magnetic moment at $\mu_0 H=5T$
decreases more drastically from $0.9\mu_B$ for x = 0.4 to
$0.15\mu_B$ for x = 1. In Fig.~\ref{F3}b, a qualitative difference
in the $M(H)$ dependence can be appreciated between the
intermediate Ni concentrated samples and those with $Ni$
concentration $\geq 75\%$, see the inset of Fig.~\ref{F3}b. While
$M(H)$ shows a tendency to saturation for x = 0.4 and 0.5, those
with x = 0.75 and 1.0 still show a positive curvature with
significantly lower values of $M$. This behavior indicates that
much larger magnetic field is required to saturate the
magnetization approaching the Ce$_2$Ni$_2$Sn stoichiometric limit,
in agreement with the rapid reduction of the Ce magnetic moment.

\begin{figure}
\begin{center}
\includegraphics[angle=0,width= 0.45 \textwidth] {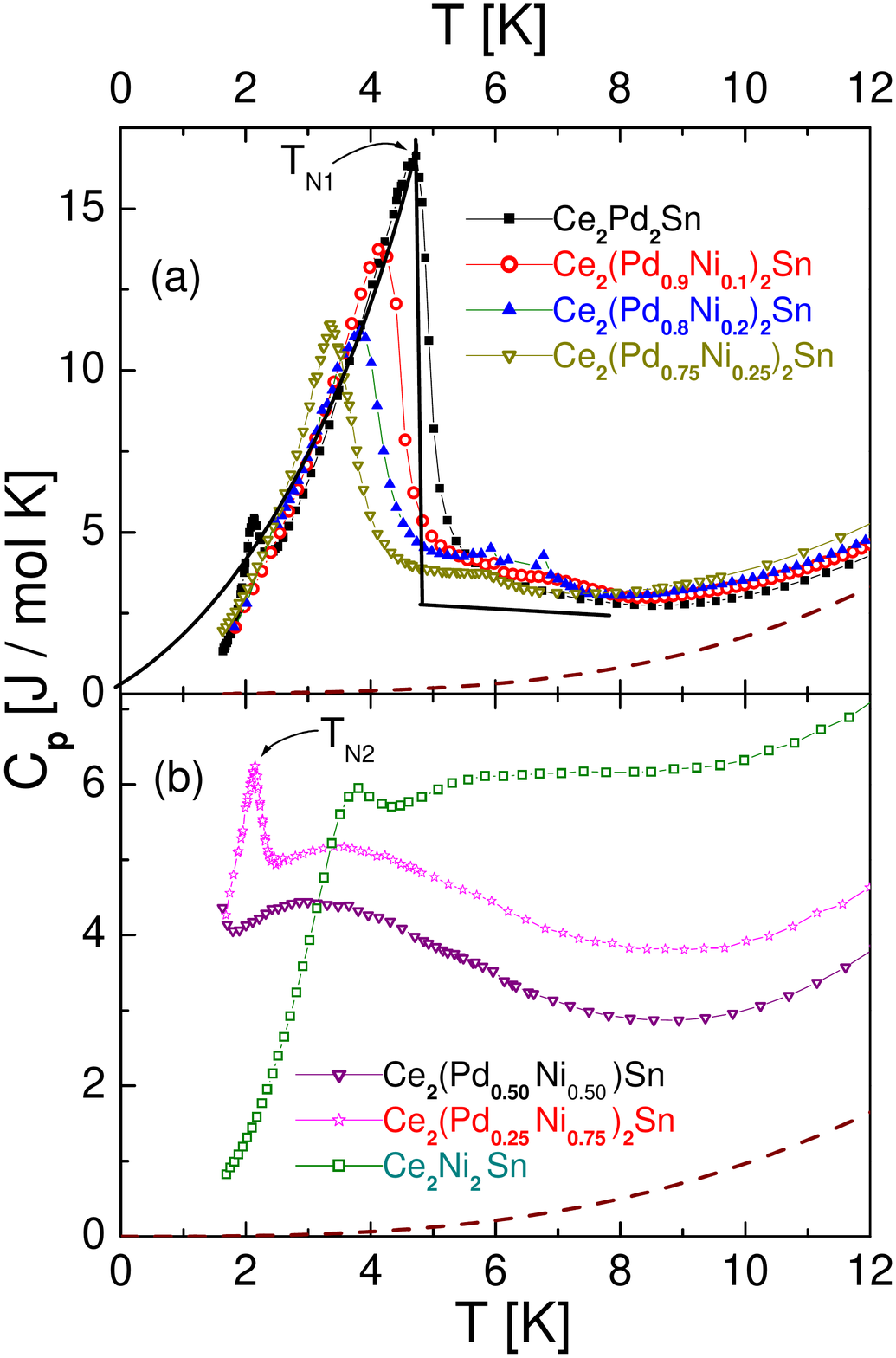}
%{/home/jsereni/papers/publicaz02/pdnial3/textos/F1latpar}
\end{center}
\caption{(Color online) Thermal dependence of specific heat $C_P)$
for (a) Pd-rich and (b) Ni-rich phases. Dashed curves represent
the respective phonon contributions extracted from La compounds.
(a) Continuous curve represents the model calculation from
$T_K/T_{N1}$ for x = 0, see the text.}
 \label{F4}
\end{figure}

\subsection{Specific heat}

The low temperature specific heat $C_P$ measurements confirm the
magnetic differences between the Ce-GS in both phases. Starting
from Ce$_2$Pd$_2$Sn which presents a double magnetic transition
\cite{Braghta}, the decrease of the AF (upper) one from
$T_{N1}=4.8$\,K to 3.3\,K for Ce$_2$(Pd$_{0.75}$Ni$_{0.25}$)$_2$Sn
can be seen in Fig.~\ref{F4}a. A concomitant decrease of the
specific heat jump at $T=T_{N1}$ is observed. On the contrary, the
FM (lower) transition at $T_C=2.2$\,K for x = 0 vanishes upon
doping.

The $C_P(T)$ tail above $T_{N1}$ reveals the presence of
increasing magnetic correlations related to the formation of
magnetic dimers between Ce-next-neighbors \cite{Ce2Pd2Sn}. A small
hump at around 6\,K can be attributed to a little amount of
foreign phase, probably CePd$_{1-x}$Ni$_x$, which orders around
that temperature \cite{CePdNi} or to the contribution of some
traces of Ce$_2$O$_3$.

The specific heat of the Ni-rich phase shows a completely
different behavior since a weak transition at $T_{N2}$, which
arises with Ni concentration, is followed by a broad maximum, see
Fig.~\ref{F4}b. Contrary to the type of transition observed in
Fig.~\ref{F4}a, the magnetic transition in Ce$_2$Ni$_2$Sn suggests
that the ordered phase is formed as a condensation of states which
may have itinerant character above $T_{N2}$. In both 4a and b
figures, the phonon references are represented by the respective
La compounds, whose contributions are subtracted in order to
obtain the magnetic $C_m(T)$ contribution.

\section{Discussion}

The temperature dependence of the magnetic susceptibility at high
temperature (c.f. $T>100$\,K) is the characteristic of a Ce system
where the six fold levels of the $J = 5/2$ Hund's rule
configuration are split by the effect of the electric crystal
field CFE. For stoichiometric Ce$_2$Pd$_2$Sn, the first excited
doublet was evaluated at $\Delta_1 \approx 50$\,K \cite{Ce2Pd2Sn}.
Such splitting largely exceeds the Kondo temperature and therefore
the low temperature magnetic properties on the Pd-rich side can be
only attributed to the Kramer's doublet GS. Such is not the case
for the Ni-rich phase, where the observed $\theta_P\propto T_K$
temperature is much larger. In this case, the possibility of an
overlap between the ground state and the first exited CFE doublet
has to be considered because $\theta_P$ and $\Delta_1$ become
comparable.

The low temperature magnetization curves clearly indicate that in
the Pd-rich phase the Ce magnetic moment behaves as strongly
localized, whereas on the Ni-rich side it rapidly decreases as the
Ni content increases. Noteworthy, the $M(H)$ curve of the x = 0.4
sample reaches a relatively high value $\mu = 0.9 \mu_B$ at 5T for
the 1.8\,K isotherm. This value is larger than the one of the x =
0.25 sample despite its $M(H)$ saturates at quite low applied
field $\mu_0 H < 1$\,T. This behavior is in line with the
localized character of the Ce-4f moment in the Pd-rich phase,
whereas on the Ni-rich one it corresponds to an itinerant type of
magnetism, which in some Ce compounds may undergo a metamagnetic
transition at very high field \cite{Micha}.

\subsection{Pd-rich side}

As mentioned before, the thermal variation of $C_P$ for $x < 0.3$
is characterized by a well defined jump $\Delta C_P$ at $T=T_{N1}$
which, for x = 0, peaks at $\Delta C_P =8.3$\,J/KmolCe instead of
$\Delta C_P(T_K=0)= 12.5$\,J/KmolCe, as expected for a S = 1/2
two-level system. By increasing $Ni$ concentration, both $\Delta
C_P(T_{N1})$ and $T_{N1}$ decrease (see Fig.~\ref{F4}a) which
indicates a moderate growing of hybridization effects driving a
progressive increase of $T_K(x)$.

\begin{figure}
\begin{center}
\includegraphics[angle=0,width= 0.45 \textwidth] {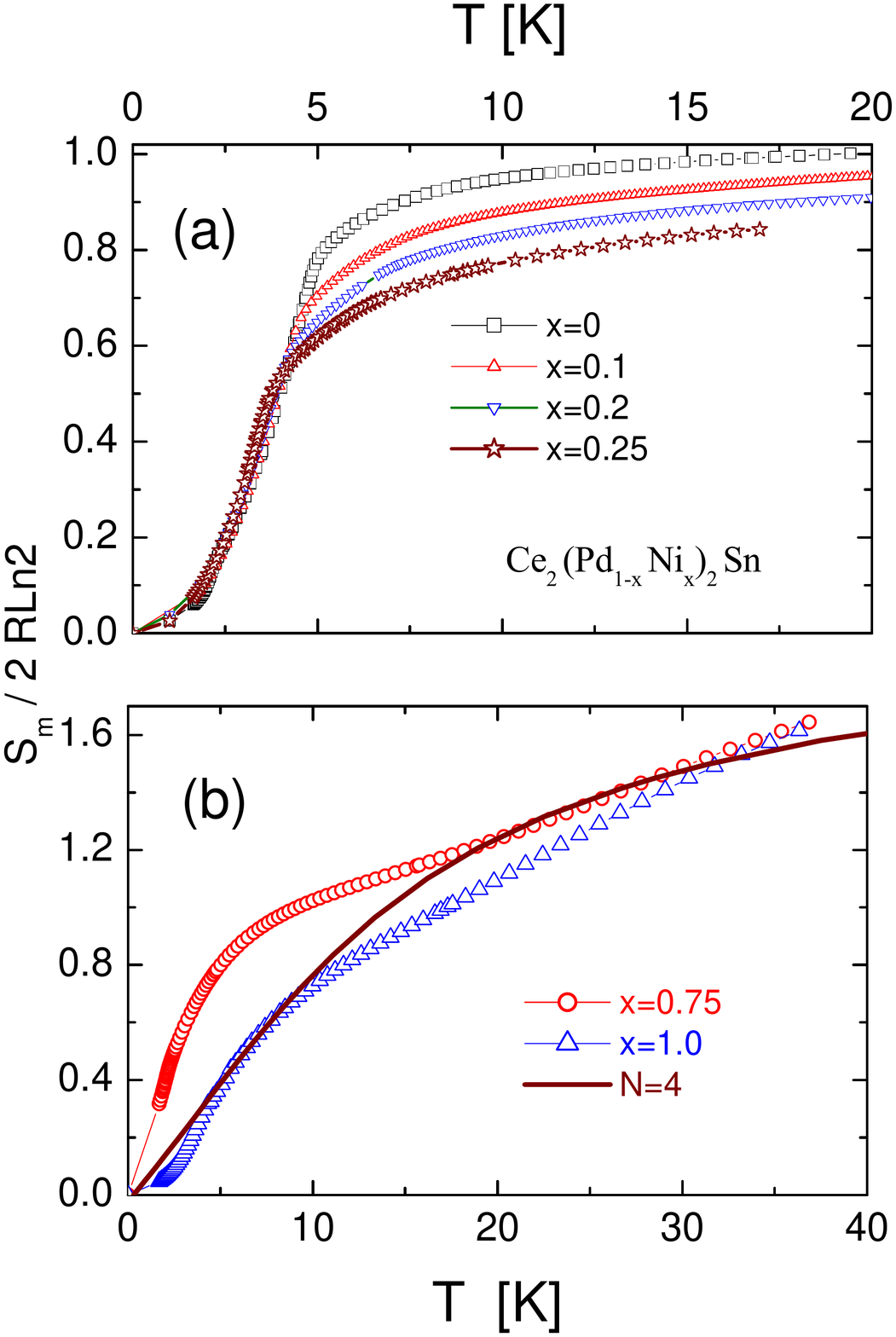}
%{/home/jsereni/papers/publicaz02/pdnial3/textos/F1latpar}
\end{center}
\caption{(Color online) Temperature dependence of the entropy. a)
for $x\leq 0.25$ up to $T=20$\,K and b) $x \geq 0.75$ up to
$T=40$\,K. Continuous curve: theoretical prediction \cite{Coqblin}
for a N=4 GS with characteristic temperature $T_o=30$\,K.}
 \label{F5}
\end{figure}

In order to evaluate the Kondo temperature $T_K(x)$ variation
within this low Ni concentration range, we have applied a resonant
level model based on molecular field calculations for spin 1/2
\cite{Braghta} to describe the $\Delta C_P$ as a function of the
$T_K/T_{N1}$ ratio. From the $\Delta C_P(x=0)/\Delta C_P(T_K=0)$
ratio a $T_K/T_{N1}=0.62$ value is obtained for the x = 0 sample,
see continuous curve in Fig.~\ref{F4}a. This value corresponds to
$T_K\approx 3$\,K, which rises up to 4.5\,K for x = 0.25 (not
shown). In spite of this increase, it is clear that $T_{N1}$ and
$T_K$ energy scales are comparable within this concentration
range.

In Fig.~\ref{F5}, we show the thermal evolution of the magnetic
entropy computed as $S_{m}=\int C_m/T dT$. Notice that the entropy
is evaluated in 'per Ce-atom' units. The behavior shown by the
alloys belonging to the Pd-rich side (see Fig.~\ref{F6}a) is the
expected for well localized 4f-moments. The plateau of $S_m(T)$
around $T=20$\,K confirms that the doublet GS is completely
occupied at that temperature and that the first excited CFE
doublet does not contribute to the magnetic GS properties. By
applying the Desgranges-Schotte \cite{Desgr} criterion of
$S_{mag}(T = T_K) \simeq 2/3 R\ln2$ for single Kondo atoms, the
extracted values of $T_K(x)$ variation nicely coincide with those
extracted from the mentioned analysis of the $\Delta C_P(T_K/T_N)$
jump.

\subsection{Ni-rich side}

\begin{figure}
\begin{center}
\includegraphics[angle=0,width= 0.5 \textwidth] {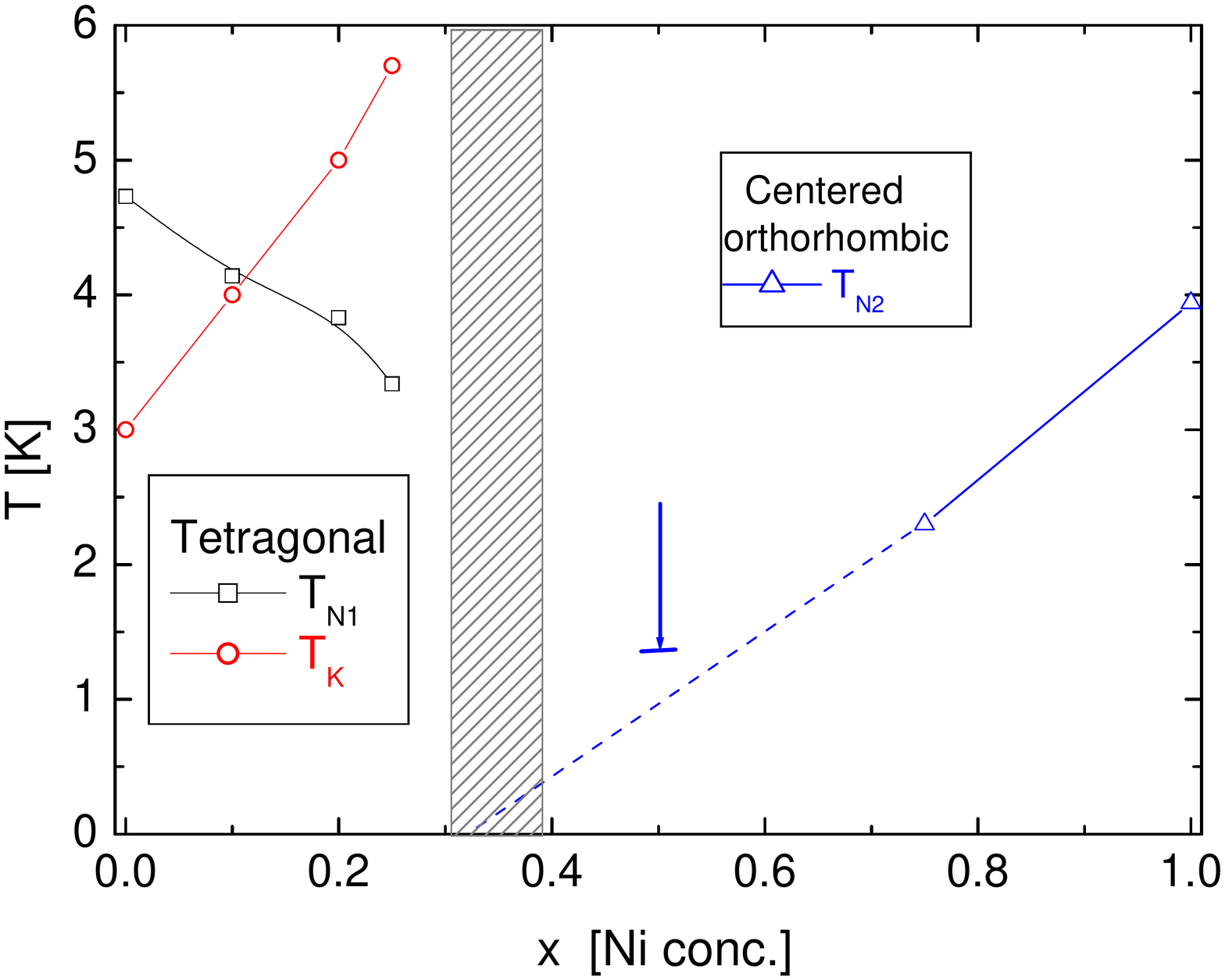}
%{/home/jsereni/papers/publicaz02/pdnial3/textos/F1latpar}
\end{center}
\caption{(Color online) Magnetic phase diagram of the
Ce$_2$(Pd$_{1-x}$Ni$_x$)$_2$Sn series including both structural
phases. The arrow indicates the lower limit of the experimental
specific heat measurement for $x = 0.5$ and the dashed line the
proposed extrapolation to $T_{N2} = 0$.}
 \label{F7}
\end{figure}

The peculiar behavior of the specific heat on the Ni-rich side
merits a deeper analysis because, as shown in Fig.~\ref{F4}b, a
weak transition at $T=T_{N2}$ is followed by a very broad maximum
at higher temperature. This unexpected thermal dependence can be
explained once the onset of $T_{N2}$ is taken into account. By
extrapolating this order temperature to zero, i.e. $T_{N2} \to 0$
decreasing Ni content as depicted in the phase diagram of
Fig.~\ref{F7}, the presence of a Quantum Critical Point QCP
\cite{HvL} can be expected at $x_{cr}= 0.35\pm 0.05$.

\begin{figure}
\begin{center}
\includegraphics[angle=0,width= 0.5 \textwidth] {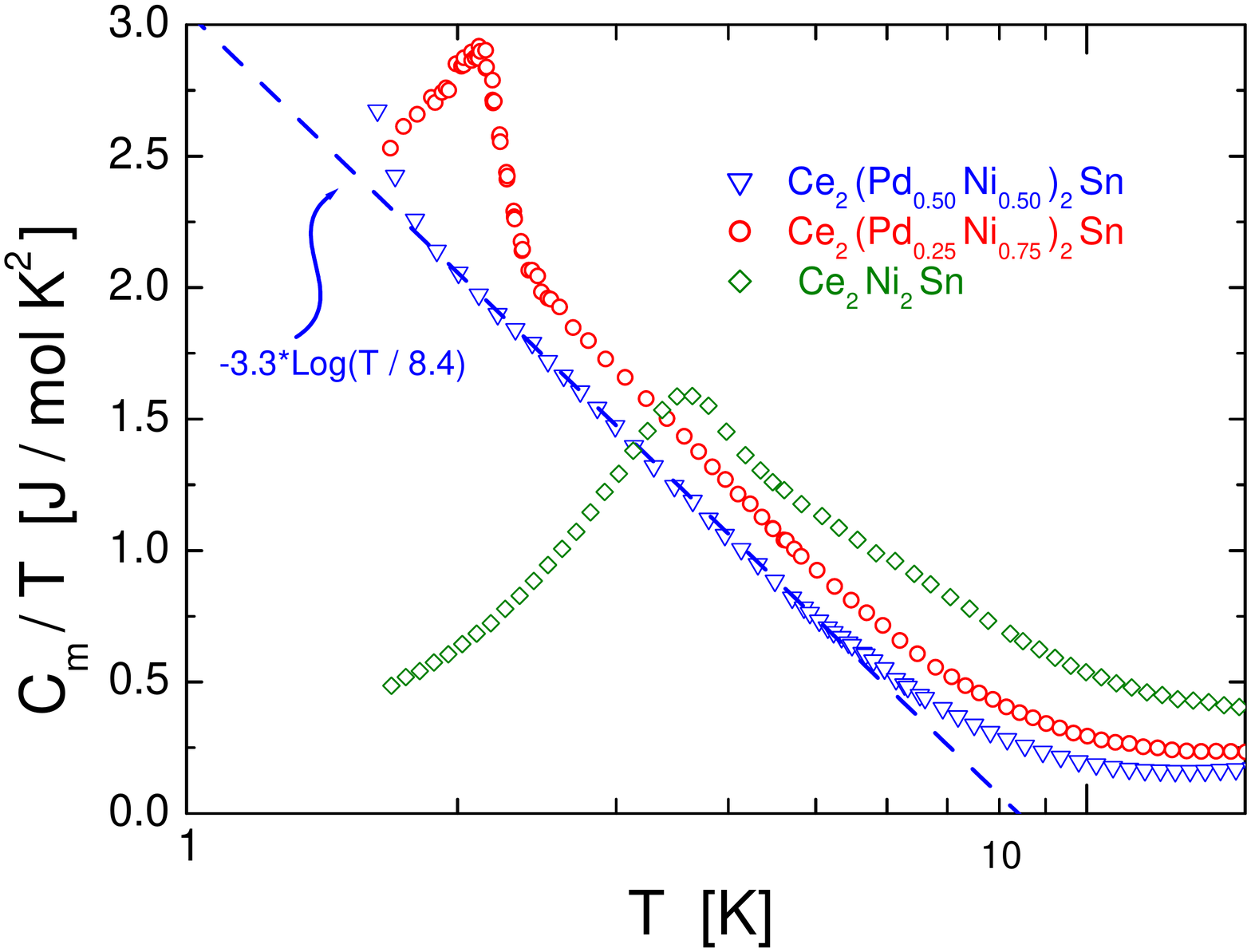}
%{/home/jsereni/papers/publicaz02/pdnial3/textos/F1latpar}
\end{center}
\caption{(Color online) Logarithmic dependence of the magnetic
contribution to specific heat divided temperature for the Ni-rich
samples. Dashed line represents the computed logarithmic reference
dependence for sample x = 0.5.}
 \label{F6}
\end{figure}

It is well known that the existence of a QCP affects the thermal
properties due to the associated low energy quantum fluctuations
which induce a peculiar thermal dependencies in the physical
parameters known as non-Fermi-liquid behavior \cite{Stewart}. One
of the theoretical predictions widely observed experimentally is a
$C_m(T)/T \propto - Log(T/T^*)$ dependence, where $T^*$ is a
characteristic energy scale describing the thermal extension of
the $C_m(T)/T$ tail. Such a behavior is observed for the x = 0.5
and 0.75 samples (with respective $T_{N2} < 1.5$ and = 2.14\,K) as
depicted in Fig.~\ref{F6}. In Ce$_2$Ni$_2$Sn that temperature
dependence above the transition is practically dominated by
classical magnetic correlations because at $T_{N2} = 3.8$\,K
quantum fluctuations are overcome by classical thermal
fluctuations \cite{JLTP}

There is an apparent contradiction within the magnetic behavior
observed in these Ni-rich samples. While their $M$(2\,K, 5\,T)
values rapidly decrease with increasing Ni content (see the inset
in Fig.~\ref{F3}b), the mentioned weak AF transition temperature
and the $\Delta C_m(T_{N2})$ jump increase from $T_{N2}\approx
1.5$\,K up to $T_{N2}\approx 3$\,K at $x=1$. This complex behavior
can be understood by comparing $T_K$ and the splitting of the
first exited CFE doublet $\Delta_1$.

According to $\theta_P(x)$ values within this Ni-rich
concentration region, an enhancement of the Kondo screening (i.e.
$T_K$) is expected respect to the tetragonal phase. Within this
context, $T_K(x)$ becomes comparable to $\Delta_1$. Unfortunately,
a direct and independent evaluation of these parameters is no
possible. However, the analysis of the thermal variation of the
magnetic contribution to the entropy provides significant
information about the broadened levels distribution.

Since the specific heat of the samples with higher Ni content
(i.e. x = 0.75 and 1.0) was measured up to 40\,K, one can trace
$S_m(T)$ up to quite high temperature, as it is shown in
Fig.~\ref{F5}b. In that figure it can be seen how $S_m(T)$
overcomes the $2 R Ln2$ value already at $T\approx 9$\,K and
continuous to grow up to $S_m =0.8 * (2RLn4)$ at T = 40\,K.
Similar behavior is observed in the $x=1$ sample (i.e.
Ce$_2$Ni$_2$Sn) but with $S_(T)$ increasing in a more monotonous
manner. Compared with the $S_m(T)$ behavior of a sharp energy
distribution of levels, provided by the results depicted in
Fig.~\ref{F5}a, it becomes evident an overlap of the side levels
of the ground and first excited CFE levels.

This is the key to understand the complex magnetic behavior of
this Ni-rich alloys which show a decrease of the magnetic moment
with a simultaneous increase of a weak $T_{N2}$. The former effect
is driven by the increase of the $4f$ state hybridization whereas
the latter is driven by the increase of the GS density of states
due to the arising contribution of the low energy levels of the
CFE excited state. Such an enhancement of the low energy density
of states is supported by the increasing $C_m/T$ contribution to
the specific heat above $T_{N2}$ deduced from Fig.~\ref{F6}, which
increases from 0.2\,J/molK$^2$ for x = 0.5 up to 0.45\,J/molK$^2$
for x=1 around $T\approx 15$\,K. Low effective moments and
relatively high ordering temperatures are typical of itinerant
magnetic systems \cite{DeLong}, provided that the density of
states is large enough. This scenario is favored by the reduction
of the Ce-Ce spacings decrease driven by the reduction of $a$ and
$b$ lattice parameters.

Within a simplified picture, such would be the expected way to
undergo from a N = 2 degenerated GS to a N = 4 degenerated GS in a
continuous way. In Fig.~\ref{F5}b we include a theoretical
prediction \cite{Coqblin} for a N=4 degenerated state with a
characteristic temperature $T_o=30$\,K. This comparison clearly
indicates that, despite of the energy levels contribution of the
first excited CFE state, the N=4 GS is not reached in this system.
Similar situation was observed in other strongly anisotropic Ce
systems such as CeTiGe \cite{Micha}.

The low temperature magnetic properties are resumed in the
magnetic phase diagram presented in Fig.~\ref{F7}. On the Pd-rich
side the comparable magnetic and Kondo energy scales can be
appreciated whereas on the Ni-rich side the extrapolation of
$T_{N2}\to 0$ is evaluated at $x_{cr} =0.35\pm 0.05$.

\section{Conclusions}

The substitution of Pd atoms by isoelectronic Ni leads to a change
in the crystallographic structure from a tetragonal lattice for $x
\leq 0.3$ to a centered orthorhombic lattice for $x \geq 0.4$
after about 1\% of volume collapse at the structural modification.

The volume contraction in the Pd-rich side induces a moderate
increase of the Kondo temperature whose value is comparable to
$T_{N1}$. This induce a decrease of $T_{N1}$ until the limit of
this phase stability. On the Ni-rich side the system behaves as a
weakly magnetic heavy fermion with an enhanced degenerate ground
state. Notably, an incipient magnetic transition $T_{N1}$ arises
in the Ni-rich side favored by the increasing low energy density
of states whereas the specific heat shows a $C_m/T \propto -
Log(T/T^*)$ dependence which is the finger print for a quantum
critical point extrapolated at $x_{cr} = 0.35$. Lower $C_m(T)$
measurements are required to very this unexpected behavior
occurring in an itinerant magnetic medium. Due to the similar
$T_K$ and CFE splitting energy scales, the ground and the first
CFE excited states start to overlap without reaching a four fold
degenerated GS.

\end{document}